\documentclass[a4paper,12pt]{article}
\usepackage{amsmath,amssymb,amsfonts,amsthm}
\newtheorem{theorem}{Theorem}[section]
\newtheorem{lemma}[theorem]{Lemma}

\newtheorem{corollary}[theorem]{Corollary}

\newcommand{\ck}{l}

\newcommand{\ek}{e}

\newcommand{\ckh}{{\mathbf l}}

\newcommand{\ekh}{{\mathbf e}}

\newcommand{\Rn}{\mathbb{R}^n}

\newcommand{\lt}{L^{2}(\Omega)}

\newcommand{\hk}[1]{H^{#1}(\Omega)}

\newcommand{\hfk}[2]{||#1||_{H^{#2}(\Omega)}}

\newcommand{\dv}{\, d\mu}

\newcommand{\dvh}{\, d\mu_h}

\title{Generalized  Korn's inequality and conformal Killing vectors}
\author{Sergio Dain\\
  Max-Planck-Institut f\"ur Gravitationsphysik\\
  Am M\"uhlenberg 1\\
  14476 Golm\\
  Germany}
  
\begin{document}
\maketitle 

\begin{abstract}

  Korn's inequality plays an important role in linear elasticity theory.  This
  inequality bounds the norm of the derivatives of the displacement vector by
  the norm of the linearized strain tensor. The kernel of the linearized strain
  tensor are the infinitesimal rigid-body translations and rotations (Killing
  vectors). We generalize this inequality by replacing the linearized strain
  tensor by its trace free part. That is, we obtain a stronger inequality in
  which the kernel of the relevant operator are the conformal Killing vectors.
  The new inequality  has applications in General Relativity. 

\end{abstract}

\section{Introduction} 
Let $\Omega$ be a domain in $\Rn$, with
$n\geq 2$ and let $u^i$ be a vector field in $\Omega$, with $i=1,\cdots,n$. 
We denote the Euclidean inner product by $u_iu^i$, where the summation
convention with respect to repeated indices is used and the indices are moved
with the Kronecker delta $\delta_{ij}$ (i.e; $u_j=\delta_{ij}u^i$). 
Let  $\hk{1}$ be the standard Sobolev space of vectors fields with norm
\begin{equation}
  \hfk{u}{1}= \left( \int_{\Omega} u_iu^i \dv  \right)^{1/2}+\left(
  \int_{\Omega} \partial_iu_j\partial^iu^j \dv \right)^{1/2},  
  \label{eq:sobolev}
\end{equation}
where $\partial_i$ denotes partial derivative with respect to the coordinate
$x_i$ and $\dv$ is the Euclidean volume element. 

For all functions $u\in \hk{1}$, there
exists a constant $C$, independent on $u$,  such that the
following inequality holds
\begin{equation}  
 \label{eq:korn} 
 \hfk{u}{1}^2  \leq C
\int_{\Omega} \left( u^iu_i + \ek_{ij}(u)\ek^{ij}(u) \right) \dv,
\end{equation}
where
\begin{equation} 
 \label{eq:ek} 
\ek_{ij}(u)= \frac{1}{2}\left( \partial_i u_j+\partial_j u_i \right).
\end{equation} 
Inequality \eqref{eq:korn} is known as Korn's inequality.  This inequality  has
a long history. It  plays a central role in elasticity, see
the review  \cite{Horgan95} and also the recent articles  \cite{Chen02},
\cite{Tiero99} and \cite{Ciarlet04}. 

In linear elasticity  $u^i$ is the displacement vector and the tensor
\eqref{eq:ek} is known as the linearized strain tensor. The solutions $u^i$ of
$\ek_{ij}(u)=0$ are the infinitesimal generator of rigid-body rotations and
translations  (Killing vectors of the flat metric), which are precisely the
only (infinitesimal) displacements which do not change the shape of the body.  The
dimension of the  kernel is  $n(n+1)/2$.

The energy of the elastic body is given by  (see, for example, \cite{Necas80})
\begin{equation}  
  \label{eq:energy}
  E(u)= \int_{\Omega}\left( c^{ijkl}\ek_{ij}(u)\ek_{kl}(u) - F^iu_i  \right)\dv, 
\end{equation} 
where $F^i$ is the external force and $c_{ijkl}$ are certain  bounded functions
which depend on the particular material. They satisfy
$c_{ijkl}=c_{klij}$, $c_{ijkl}=c_{jikl}$, $c_{ijlk}=c_{ijkl}$ and  
 the positivity condition
\begin{equation} 
  \label{eq:ener-pos}
 c^{ijkl}\ek_{ij}(u)\ek_{kl}(u)\geq C_0 \ek_{ij}(u)\ek^{ij}(u), 
\end{equation}
for some constant  $C_0>0$. From \eqref{eq:energy} and 
\eqref{eq:ener-pos} we deduce that if $F^i=0$ then  $E(u)=0 \iff e_{ij}(u)=0$. That is, in
absence of external forces,  the
energy of a displacement $u^i$ is zero if and only if $u^i$ is a rigid-body
translation or rotation, in accordance with physical intuition.     

The pure traction problem of linear elasticity (i.e; where the forces at the
boundary are prescribed) consists in finding a
displacement $u^i\in \hk{1}$ that minimize \eqref{eq:energy}. Korn's inequality
is used to prove that the functional \eqref{eq:energy} is coercive, existence
of weak solutions then follows by the Lax-Milgram theorem.

We define the operator $\ck_{ij}(u)$ as the trace free part of
$\ek_{ij}(u)$
 \begin{equation}  
 \label{eq:ck} 
 \ck_{ij}(u)=\ek_{ij}(u)-\frac{1}{n}\ek(u) \delta_{ij}, 
\end{equation}  
where 
\begin{equation} 
        \ek(u)=\delta^{ij}\ek_{ij}(u)=\partial^iu_i.
        \label{eq:tek}
\end{equation} 
The following is the main result of this article. 
\begin{theorem}
        \label{t1}
        Let $\Omega$ be a Lipschitz domain in $\Rn$, with $n\geq 3$. Then, there exists
        a constant $C$, independent of $u$,  such that
\begin{equation}  
 \label{eq:kkorn} 
  \hfk{u}{1}^2  \leq C  
  \int_{\Omega}\left( u^iu_i + \ck_{ij}(u)\ck^{ij}(u) \right)\dv, \text{ for all } u\in\hk{1}.
\end{equation}

\end{theorem}
For a definition of  Lipschitz domains see \cite{Adams} and
\cite{Necas80}. These domains include domains
with corners like cubic domains.

Inequality \eqref{eq:kkorn} implies inequality
\eqref{eq:korn} since we have
\begin{equation}  
 \label{eq:ckek} 
\ck_{ij}(u)\ck^{ij}(u)=\ek_{ij}(u)\ek^{ij}(u)-\frac{1}{n}(\ek(u))^2\leq
\ek_{ij}(u)\ek^{ij}(u).
\end{equation} 
The kernel $\ck_{ij}(u)=0$ is given by the conformal Killing vectors, which
include the Killing vectors and also the dilatations and special
conformal transformations given by 
\begin{equation}
  \label{eq:flatcon}
ax^i, \quad k^j\,(2\,x_j\,x^i - \delta_j\,^i\,x_l\,x^l),
\end{equation}
where  $k^i$ and $a$  are arbitrary constants. If $n\geq 3$ then the
dimension of the kernel is $(n+1)(n+2)/2$. 

For vectors $u^i$ which vanish at the boundary $\partial\Omega$ Korn's
inequality \eqref{eq:korn}  follows easily by integration  by parts (this is
known in the literature as Korn's inequality in the first case).  The same
argument applies to the
inequality \eqref{eq:kkorn}: if we assume that $u^i=0$ on $\partial\Omega$,
then for $n\geq 2$ we have
\begin{align}
 2 \int_{\Omega} \ck_{ij}(u)\ck^{ij}(u)\dv & =   \int_{\Omega}
 \partial_i u_j \partial^i u^j\dv +\frac{(n-2)}{n}  \int_{\Omega}(\partial^iu_i)^2 \dv\\
& \geq   \int_{\Omega}
 \partial_i u_j \partial^i u^j\dv 
  \label{eq:div}
\end{align}

Note that we can replace the operator $\ck$ in \eqref{eq:kkorn} by 
\begin{equation}
  \label{eq:2}
  \ck'_{ij}(u)=\ek_{ij}(u)-\alpha\delta_{ij} \ek(u),
\end{equation}
where $\alpha$ is an arbitrary real number, because we have the inequality 
\begin{equation}
  \label{eq:3}
  \ck_{ij}(u)\ck^{ij}(u)\leq  \ck'_{ij}(u) {\ck'}^{ij}(u)=
  \ck_{ij}(u)\ck^{ij}(u)+\frac{(\ek(u)(1-n\alpha))^2}{n} 
\end{equation}

The case $n=2$ is special. 
In this case, equations $\ck(u)_{ij}=0$ are given by 
\begin{equation}
  \label{eq:cr}
 \partial_1u_1-\partial_2u_2=0, \quad  \partial_1u_2-\partial_2u_1=0.
\end{equation}
These equations  are the Cauchy-Riemann equations for the complex function
$F=u_1 + i u_2$. That is, every analytical function $F$ provides  a solution of
$\ck(u)_{ij}=0$. Then, for $n=2$, the kernel of the operator $\ck$ is infinite
dimensional. This implies that theorem  \ref{t1} \emph{does not hold} for $n=2$ as we
will see. Note that Korn's inequality \eqref{eq:korn} holds in this case
and also inequality \eqref{eq:div} for vectors $u^i$ which vanish at the
boundary. 

To prove that theorem \ref{t1} is not valid in two dimensions we argue by
contradiction. Let us assume that inequality \eqref{eq:kkorn} holds. Then,
following the same argument used in \cite{Chen02} to prove that Korn's
inequality implies that the kernel of $\ek$ is finite dimensional, we conclude
that the kernel of $\ck$ is finite dimensional. This provides the required
contradiction for $n=2$. For $n\geq 3$ this is an alternative proof of the
above mentioned fact that the space of conformal Killing vectors is finite
dimensional.

The operators $\ek_{ij}(u)$ and $\ck_{ij}(u)$ have a natural generalization for
Riemannian manifolds.  Let $M$ be a
Riemannian manifold with metric $h_{ij}$ and covariant derivative $D_i$. Then, the
operators $\ek$ and $\ck$ generalize to
\begin{equation} 
 \label{eq:ekh} 
\ekh_{ij}(u) = \frac{1}{2}\left( D_i u_j+D_j u_i\right)
 \end{equation}
and
\begin{equation}
  \label{eq:ckh}
\ckh_{ij}(u)= \ekh_{ij}(u)-\frac{1}{n}\ekh(u)h_{ij},  
\end{equation}
where
\begin{equation}  
        \ekh(u)=h^{ij}\ekh_{ij}=D_iu^i,
        \label{eq:tekh}
\end{equation}
and the indices are moved with the metric $h_{ij}$ and its inverse $h^{ij}$
(i.e; $u_i=h_{ij}u^j$).  

The operator $\ckh$ is conformal invariant in the following sense. If $\tilde
h_{ij}=e^{2f}h_{ij}$ is a metric conformal to $h_{ij}$ (where $f$ is an
arbitrary function) and $\tilde \ckh$ is the corresponding operator, then we
have the relation
\begin{equation}
  \label{eq:crckh}
  \tilde \ckh(\tilde u)_{ij}= e^{2f} \ckh(u)_{ij},
\end{equation}
where $\tilde u^i=u^i$ and $\tilde u_i=e^{2f}u_i$ (in equation
\eqref{eq:crckh}, indices of quantities with
tilde are moved with the metric $\tilde h_{ij}$ and its inverse). 

In \cite{Chen02} a Riemannian version of Korn's
inequality was proved. Using the same arguments and theorem \ref{t1} the
following result follows.  
\begin{corollary}
Let $\Omega\subset M$ be an open set with Lipschitz boundary and assume that
the metric $h_{ij}$ is in $C^1(M)$. Then, there is a
positive constant $C$ such that 
\begin{equation}  
 \label{eq:rkkorn} 
 \int_{\Omega} D_i u_j uD^i u^j \dvh   \leq C
\int_{\Omega}u^iu_i + \ckh_{ij}(u)\ckh^{ij}(u)\dvh,\text{ for all } u\in\hk{1},  
\end{equation} 
where $\dvh$ is the volume element of the metric $h_{ij}$.
\end{corollary}

Inequality \eqref{eq:rkkorn} has applications in General Relativity.  In the
Cauchy formulation of the theory, the initial data have to satisfy the so
called constraint equations on a Riemannian manifold (see the recent review
article \cite{Bartnik:2002cw} and reference therein). Solutions of one of these
equations (the ``momentum constraint'') can be obtained as a solution of a
variational problem for the following energy under suitable boundary conditions
for $u^i$
\begin{equation} 
  \label{eq:energyc}
E'(u)= \int_{\Omega}\left ( \ckh_{ij}(u)\ckh^{ij}(u)- Q^iu_i\right) \dvh,
\end{equation}
where $Q^i$ is a given vector. 
The energy \eqref{eq:energyc} has similar form to the elastic energy
\eqref{eq:energy}, the difference is that the strain tensor $\ekh_{ij}$ is replaced by
$\ckh_{ij}$. For black holes, the boundary conditions for $u^i$
are  analogous to the pure traction problem of linear elasticity. That is, the
solution is the infimum of $E'(u)$  for all $u\in \hk{1}$ (see \cite{Dain03},
\cite{Dain05}).
As in the case of elasticity, inequality \eqref{eq:kkorn} is used to prove that $E'(u)$ is coercive in $u\in
\hk{1}$, and then the existence of solution follows by the Lax-Milgram theorem
(see  \cite{Dain03}, \cite{Dain05} for details).

In elasticity, the energy \eqref{eq:energyc} has no direct physical meaning
since in absence of external forces it is zero not only for rigid-body
displacement but also for dilatations. That is, the  ``bulk modulus''
coefficient of the material is equal to zero; no elastic material has this
property. On the other hand, the constraint equations of General Relativity are
conformal invariant (see \cite{Bartnik:2002cw}), this is why  $\ckh_{ij}$  and
not $\ekh_{ij}$ appears in \eqref{eq:energyc}.

\section{Proof of Theorem \ref{t1}}

The strategy of the proof follows the proof of Korn's inequality given in
\cite{Necas67} and \cite{Duvaut76}. The main tool is the following
remarkably lemma proved in \cite{Necas67} (see also \cite{McLean00}).

\begin{lemma}
  \label{necas}
Let $\Omega$ be  a Lipschitz domain and let $u$ be a distribution on $\Omega$
such that $u\in \hk{-p-q}$ and $\partial^\alpha u \in \hk{-p-q}$, $|\alpha|
\leq q$, for some integers  $p\geq 0$ and
$q\geq1$. Then $u\in \hk{-p}$. 
\end{lemma}
 For the definition of Sobolev spaces with negative exponents see
 \cite{Lions72}.  We will use the standard notation $\hk{0}=\lt$

This lemma is a generalization of Theorem 3.2, Chapter III, page 111, in
\cite{Duvaut76} (see also Remark 3.1 on page 112 and the Comments (section 8)
on page 196 in the same chapter) where the case $q=0$, $p=1$ is proved. This
particular 
case is enough for proving Korn's inequality. However, for the inequality
\eqref{eq:kkorn} we need to take one more derivative, and hence we will use
lemma \ref{necas} for $q=p=1$. 

\begin{proof}
Following \cite{Duvaut76}, we divide the proof in two steps. 

\emph{Step 1.} Using lemma \ref{necas}, we will prove that
$u_{i},\ck_{ij}(u)\in \lt$ implies $u_i\in \hk{1}$. 

We have the following identity
 \begin{equation}
        \partial_k \partial_j u_i = \partial_j \ek_{ik}(u)+  \partial_k \ek_{ij}(u)-
\partial_i \ek_{jk}(u).
\label{eq:idenek}
\end{equation}
From this we deduce
\begin{equation}
        \partial_k \partial_j u_i = \partial_j \ck_{ik}(u)+  \partial_k \ck_{ij}(u)-
        \partial_i \ck_{jk}(u)+\frac{1}{n} \left(- \partial_j\ek(u) \delta_{ik}- \partial_k\ek(u) \delta_{ij}+\partial_i\ek(u) \delta_{jk}\right).
\label{eq:idenck}
\end{equation}
Taking a derivative $\partial^k$ of equation \eqref{eq:idenck} we obtain
\begin{equation}
        \partial_j \Delta u_i=\partial_j
        \partial^k\ck_{ik}+\Delta\ck_{ij}-\partial_i
        \partial^k\ck_{jk}-\frac{1}{n-1}\delta_{ji}\partial^k
        \partial^f\ck_{kf},
        \label{eq:idenlap}
\end{equation}
where we have used 
\begin{equation}
  \partial^i\partial^j\ck_{ij}(u)=\frac{(n-1)}{n}\Delta \ek(u).
  \label{eq:idenlap2}
\end{equation}

By hypothesis we have $\ck_{ij}(u)\in\lt$, then the right hand side of equation 
\eqref{eq:idenlap} is in $\hk{-2}$ and hence $\partial_j \Delta u_i\in
\hk{-2}$. We use Lemma \ref{necas} for the functions $\Delta u_i$ with $p=q=1$
to conclude that $\Delta u_i\in \hk{-1}$. Then, by  the  identity
\begin{equation}
  \partial^i\ck_{ij}(u)=\frac{1}{2}\Delta u_j + \left( \frac{1}{2}-
  \frac{1}{n}\right) \partial_j\ek(u)
  \label{eq:idenck2}
\end{equation} 
we conclude that $\partial_j\ek(u)\in \hk{-1}$. 

Going back to  equation \eqref{eq:idenck} and using $\partial_j\ek(u)\in
\hk{-1}$ we conclude that  $ \partial_k \partial_j u_i\in \hk{-1}$. We apply
again Lemma \ref{necas} for the function $\partial_j u_i$ with $p=0$ and $q=1$
and we obtain  $\partial_j u_i\in \lt$, that is, $u_i\in \hk{1}$.

\emph{Step 2.} Let $H$ be the space of $u^i\in \lt$ such that $\ck_{ij}(u)\in
\lt$. $H$ is a Hilbert space  for the norm
\begin{equation}
  \label{eq:hilbertck}
  \int_{\Omega}\left( u^iu_i + \ck_{ij}(u)\ck^{ij}(u)\ \right)dv. 
\end{equation}
In Step 1 we have proved that $u\in \hk{1} \iff u\in H$. We apply the closed
graph theorem to the identity mapping from $\hk{1}$ into $H$ to obtain 
inequality \eqref{eq:kkorn}.

\end{proof}
 
The proof fails for  $n=2$ because in this case we can not
use equation \eqref{eq:idenck2} to conclude that  $\partial_j\ek(u)\in
\hk{-1}$. 

As it was mentioned  in \cite{Dain05}, an alternative proof of this
theorem, under stronger assumptions on the regularity of the boundary, can be
obtained using Proposition 12.1 of \cite{Taylor96}. 

\section*{Acknowledgments}
It is a pleasure to thank H. Friedrich for 
discussions regarding the case $n=2$.
This work has been supported by the Sonderforschungsbereich SFB/TR7
of the Deutsche Forschungsgemeinschaft.


\end{document}